\documentclass[useAMS,articles,fleqn] {mnras}
\usepackage{epsfig}
\usepackage{amsmath}
\usepackage{graphicx}
\usepackage{amssymb}

\def\Q{\ifmmode\mathcal{Q}\else$\mathcal{Q}$\fi}

\title[Structure and mass segregation in Galactic clusters]{Structure and mass segregation in Galactic stellar clusters}

\author[Dib et al.]{Sami Dib$^{1,2}$\thanks{E-mail, SD: sami.dib@gmail.com}, Stefan Schmeja$^{3,4}$, Richard J. Parker$^{5}$\thanks{Royal Society Dorothy Hodgkin Fellow}\\ 
$^{1}$Instituto de Astronom\'{i}a y Ciencias Planetarias, Universidad de Atacama, Copayapu 485, Copiap\'{o}, Chile.\\
$^{2}$Niels Bohr International Academy, Niels Bohr Institute, Blegdamsvej 17, DK-2100 Copenhagen, Denmark\\
$^{3}$Technische Informationsbibliothek, Welfengarten 1b, 30167 Hannover, Germany\\
$^{4}$Astronomisches Rechen-Institut, Zentrum f\"{u}r Astronomie der Universit\"{a}t Heidelberg, M\"{o}nchhofstra{\ss}e 12-14, 69120 Heidelberg, Germany\\ 
$^{5}$Department of Physics and Astronomy, The University of Sheffield, Hicks Building, Hounsfield Road, Sheffield S3 7RH, United Kingdom\\
}

\begin{document} 
\maketitle

\date{Received xxx; accepted yyy}

\begin{abstract} 
 
We quantify the structure of a very large number of Galactic open clusters and look for evidence of mass segregation for the most massive stars in the clusters. We characterise the structure and mass segregation ratios of 1276 clusters in the Milky Way Stellar Cluster (MWSC) catalogue containing each at least 40 stars and that are located at a distance of up to $\approx 2$ kpc from the Sun. We use an approach based on the calculation of the minimum spanning tree of the clusters, and for each one of them, we calculate the structure parameter \Q\ and the mass segregation ratio $\Lambda_{\rm MSR}$. Our findings indicate that most clusters possess a \Q\ parameter that falls in the range 0.7-0.8 and are thus neither strongly concentrated nor do they show significant substructure. Only 27\% can be considered centrally concentrated with \Q\ values $> 0.8$. Of the 1276 clusters, only 14\% show indication of significant mass segregation ($\Lambda_{\rm MSR} > 1.5$). Furthermore, no correlation is found between the structure of the clusters or the degree of mass segregation with their position in the Galaxy. A comparison of the measured \Q\ values for the young open clusters in the MWSC to N-body numerical simulations that follow the evolution of the \Q\ parameter over the first 10 Myrs of the clusters life suggests that the young clusters found in the MWSC catalogue initially possessed local mean volume densities of $\rho_{*} \approx 10-100$ M$_{\odot}$ pc$^{-3}$.
\end{abstract}
   

\section{Introduction}

Star clusters are fundamental building blocks of galactic discs and most stars, if not all, form in clusters (e.g., Carpenter 2000; Lada \& Lada 2003; Dib et al. 2011a,b; Dib 2011; Mallick et al. 2014; Dib 2014; Hony et al. 2015). The dynamics of stars in the clusters as well as the structure of clusters measured as a function of cluster age hold important clues on the processes of star formation and stellar evolution. As clusters age, the expulsion of gas by stellar feedback as well as dynamical interactions between stars and binary systems in the cluster soften its gravitational potential, leading to their expansion and to their partial or total dissolution into the field of their host galaxy (e.g., Spitzer \& Harm 1958; Parker \& Meyer 2012; Parker \& Dale 2013; Dib et al. 2011a,b; Dib 2011; Dib et al. 2013; Pfalzner \& Kaczmarek 2013; Brinkmann et al. 2017). Clusters can also be disrupted by close encounters with giant molecular clouds as they orbit the Galactic centre (e.g., Gieles et al. 2006) or by being subjected to strong tidal fields (e.g., Dalessandro et al. 2015; Zhai et al. 2017; Martinez-Medina et al. 2017). 

The initial spatial distribution of stars in young clusters may reflect the structure of the parental protostellar clump/cloud (e.g., Dib et al. 2010a; Lomax et al. 2011; Gouliermis et al. 2014; Hony et al. 2015). However, as the clusters evolve, their structure is shaped by the gravitational interactions between member stars and by tidal effects, and the structure of the clusters will reflect their dynamical evolution. Numerical simulations of star cluster formation show that clusters can build up in a hierarchical way from several sub-clusters which evolve dynamically and merge into a single, centrally concentrated cluster (e.g., Bonnell \& Bate 2006; Schmeja \& Klessen 2006; Moeckel \& Bate 2010; Allison et al. 2010; Padoan et al. 2014; Parker et al. 2014; Fujii 2015) or from the direct collapse of a single gravitationally bound clump (e.g., Banerjee \& Kroupa 2015). The latter scenario is likely to be required in order to reproduce the high star formation efficiencies and short age spreads observed in massive clusters (Dib et  al. 2013). Substructure in a fractal cluster may be erased rapidly or preserved for a longer time, depending on the stellar velocity dispersions. Results from N-body simulations (e.g, Goodwin \& Whitworth 2004) indicate that in clusters with low initial stellar velocity dispersions, the resulting collapse of the cluster tends to erase substructure to a large extent. In clusters with virial ratios\footnote{Our definition of the virial ratio is $\alpha_{\rm vir}=E_{k}/E_{grav}$, where $E_{k}$ and $E_{grav}$ are the total kinetic and potential energy, respectively. In the N-body models we compare our observation to in \S.~\ref{implicationsf}, a initial value of $\alpha_{\rm vir} < 0.3$ refers to subvirial initial conditions, whereas $\alpha_{\rm vir}=0.5$ refers to a virial case. However, since the models considered in \S.~\ref{implicationsf} have spatial and velocity substructure, a value of $\alpha_{\rm vir}=0.5$ does not necessarily imply virial equilibrium. For more discussion on this point, we refer the reader to Parker et al. (2014).} of $0.5$ or higher, however, initial substructure survives for several crossing times. Spatial substructure has been observed in clusters as old as $\approx$100 Myr (e.g, S\'{a}nchez \& Alfaro 2009). However, the structure of open clusters may also be a result of later dynamical evolution.

In many star clusters, the brightest, most massive stars are concentrated toward the centre of the cluster, which is usually attributed to mass segregation (e.g., Dib et al. 2010a; Hasan et al. 2011; Haghi et al. 2015; Sheikhi et al. 2016). Whether mass segregation occurs due to an evolutionary effect or is of primordial origin is not  yet entirely clear. In the first case, massive stars formed elsewhere in the cluster eventually sink to the cluster centre through the effects of two-body relaxation (e.g., McMillan et al. 2007; Allison et al. 2009a). This is corroborated by numerical simulations in which mass segregation occurs on timescales that are of the order of the clusters ages (e.g., Allison et al. 2010; Parker et al. 2014). In the second scenario, massive stars form preferentially in the central region of the cluster either by efficiently accreting gas due to their location at the bottom of the cluster potential well (e.g., Dib et al. 2010a) or by a coalescence process of less massive stars (Dib 2007; Dib et al. 2007a,2008a). The fact that mass segregation is also observed in young clusters (e.g., Littlefair 2003; Gouliermis et al. 2004; Stolte et al. 2006; Sharma et al. 2007; Chen et al. 2007; Gennaro et al. 2011; Pang et al. 2013; Feigelson et al. 2013; Habibi et al. 2013) might suggest that the second scenario is more likely, but the question is still under intense debate. 

In the Milky Way, several studies have explored the dependence (or lack of it) of some of the properties of open clusters with their age and surface density, such as their members richness (e.g., Tadross et al. 2002), size (Schilbach et al. 2006; Tadross 2014), galactic scale height (Buckner \& Froebrich 2014), mass segregation (Bukowiecki et al. 2012) or structure (Gregorio-Hetem et al. 2015), as well as the dependence of the cluster structure and metallicity on their position in the Galactic disk (e.g., Friel 1995; Froebrich et al. 2010; Bukowiecki et al. 2011; Tadross 2014). The Milky Way Star Cluster (MWSC) catalogue (Kharchenko et al. 2012,2013; Schmeja et al. 2014; Scholz et al. 2015; Dib et al. 2017) offers the largest homogeneous sample of Galactic open clusters, allowing us to study the spatial structure and mass segregation in a large number of clusters over a wide range of cluster ages ranging from young clusters with ages $\approx 1$ Myr to older clusters with ages of about 5 Gyr. In \S.~\ref{secdata} we briefly recount some of the characteristics of the MWSC, while in \S.~\ref{secmethods} we describe the methods we use to describe the structure and mass segregation of clusters in the catalogue. The results on the existence of correlations (or lack of it) between the structure and mass segregation levels in the clusters versus cluster properties are presented and discussed in \S.~\ref{secresults},  and in \S.~\ref{secconc}, we conclude. 

\section{Data}\label{secdata}

\begin{figure}
\begin{center}
\epsfig{figure=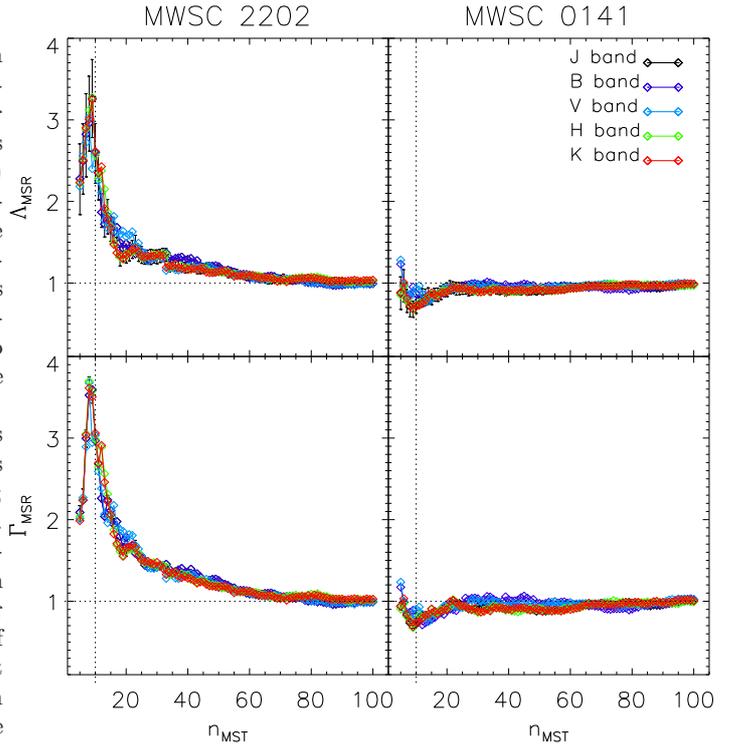,width=\columnwidth}
\end{center}
\vspace{0.5cm}
\caption{Examples of the mass segregation ratios $\Lambda_{\rm MSR}$ (Allison et al. 2009a, top row) and $\Gamma_{\rm MSR}$ (Olczak et al. 2011, bottom row) as a function of the number of stars used for computing them $n_{\rm MST}$, using the different bands available in the MWSC catalogue for two clusters. For the sake of clarity, error bars are only shown for the $J$ band measurements. The left panels display the case of a clearly mass-segregated cluster (MWSC 2202 = NGC 5460) and the right panels a cluster without any signs of mass segregation (MWSC 141 = ASCC~6, lower panel). The horizontal dotted line shows the division between non-mass-segregated and mass-segregated clusters at $\Lambda_{\rm MSR} = 1$ (or $\Gamma_{\rm MSR}=1$). The vertical line is placed at $n_{\rm MST} = 10$, the value used for comparing the mass segregation ratios in the remaining of the text.}
\label{fig1}
\end{figure}

The Milky Way Star Cluster (MWSC) catalogue (Kharchenko et al. 2012,2013) with its latest addition of predominantly old open clusters (Schmeja et al. 2014) contains 3145 confirmed Galactic open clusters, globular clusters, and compact associations. They have been analysed in a homogeneous way using 2MASS and PPMXL (R\"{o}ser et al. 2010), resulting in uniform structural, kinematic, and astrophysical data, such as radii, tidal radii, distances, ages, as well as the membership probability of stars in the cluster, among several other properties. Scholz et al. (2015) added 63 additional clusters to the catalogue, raising the total number to 3208. Apart from about 60 old (ages $\gtrsim 1$ Gyr) open clusters missing within 1 kpc of the Sun (Schmeja et al. 2014), the MWSC is complete to a distance of about 1.8 kpc. We also refer the reader to Dib et al. (2017) for further discussion on the completeness of the MWSC catalogue with respect to various implied initial cluster mass functions in the Galaxy. The MWSC contains spatial, kinematic, and photometric membership probabilities $P_{\rm s}$, $P_{\rm kin}$, $P_{JH}$, and $P_{JK}$ for each star within the cluster area. For more details on the determination of these probabilities, see Kharchenko et al. (2012). The combined membership probability is thus defined as:

\begin{equation}
P = P_{\rm s} \cdot \min(P_{\rm kin}, P_{JH}, P_{JK}).
\label{eq1}
\end{equation}

In this work, following Kharchenko et al. (2013), we consider a star to be a cluster member when it has a membership probability of $P \ge 61$\% or higher. In addition, we require that the 2MASS quality flag $Q\_flg$ = A (corresponding to a signal-to-noise ratio $S/N > 10$) in each photometric band for stars fainter than $K_s = 7$ (Kharchenko et al. 2012).

\section{Methods}\label{secmethods}

In order to study the clusters structure and mass segregation, we use two methods based on a minimum spanning tree (MST) which is the unique set of straight lines ('edges') connecting a given set of points without closed loops, such that the sum of all edge lengths is a minimum (Bor\r{u}vka 1926; Kruskal 1956; Prim 1957; Gower \& Ross 1969). These methods are detailed below.

\subsection{Structure parameter}\label{qparam}

A commonly used quantity to characterise the structure of clusters is the \Q\ parameter (Cartwright \& Whitworth 2004,2009) which is given by:
 
\begin{equation}
\Q = \frac{\bar{\ell}_{\rm MST}}{\bar{s}}.
\label{eq2}
\end{equation}

The parameter combines the normalised correlation length $\bar{s}$, i.e., the mean distance between all stars, and the normalised mean edge length $\bar{\ell}_{\rm MST}$ derived from the MST. The \Q\ parameter is used to quantify the structure of a cluster and to distinguish between clusters with a central density concentration and hierarchical clusters with a fractal substructure. Large \Q\ values ($\Q\ > 0.8$) are associated with centrally condensed clusters with radial density profiles $\rho(r) \propto r^{-\alpha}$, while small \Q\ values ($\Q\ < 0.8$) indicate clusters with a fractal substructure. \Q\ is correlated with $\alpha$ for \Q\ $>0.8$ and anticorrelated with the fractal dimension $D$ for \Q\ $< 0.8$ (Cartwright \& Whitworth 2004, in particular see Figure 5 in their paper). An interesting aspect of the \Q\ parameter is that it measures the level of substructure present in a cluster independent of the cluster density. A detailed description of the method, and in particular its implementation used in this study, is given in Schmeja \& Klessen (2006). 

\subsection{Mass segregation ratio}\label{msrparam}

Allison et al. (2009a) introduced the mass segregation ratio ($\Lambda_{\rm MSR}$) as a measure to identify and quantify mass segregation in clusters. The method is based on a calculation of the length of the MST, $l_{\rm MST}$, which measures the compactness of a given sample of vertices in the MST. The mass segregation of a cluster is measured by comparing the value of $l_{\rm MST}$ of the $n_{\rm MST}$ most massive stars, $l^{mp}_{\rm MST}$, with the average $l_{\rm MST}$ of k sets of n random stars, $\left< l^{\rm rand}_{\rm MST}\right>$. The value of $\Lambda_{\rm MSR}$ is then given by:

\begin{equation}
\Lambda_{\rm MSR}= \frac{\left<l^{\rm rand}_{\rm MST}\right>}{l^{\rm mp}_{\rm MST}}. 
\label{eq3}
 \end{equation}
 
The error on $\Lambda_{\rm MSR}$ is given by:

\begin{equation}
\Delta \Lambda_{\rm MSR}=\frac{\Delta l^{\rm rand}_{\rm MST}}{l^{\rm mp}_{\rm MST}},
 \label{eq4}
 \end{equation}
 
\noindent where $\Delta l^{\rm rand}_{\rm MST}$ is the standard deviation from the k random sets. The method has been modified by Olczak et al. (2011) by using the geometric mean rather than the arithmetic mean in order to minimise the influence of outliers. This method works by constructing the MST for the $n_{\rm MST}$ most massive stars and determining the mean edge length $\gamma_{\rm mp}$. Then, we construct the MST of the same number of randomly selected stars from the entire sample and determine the mean edge length $\gamma_{\rm rand}$. The value of the MSR following Olczak et al. (2011), $\Gamma_{\rm MSR}$, is then given by:
 
\begin{equation}
\Gamma_{\rm MSR} = \frac{\langle \gamma^{\rm rand}_{\rm MST} \rangle}{\gamma^{\rm mp}_{\rm MST}},
\label{eq5}
\end{equation}

and the associated standard deviation of $\Gamma_{\rm MSR}$ is given by:

\begin{equation}
\Delta \Gamma_{\rm MSR}=\Delta \gamma^{\rm rand}_{\rm MST}.
\label{eq6}
\end{equation}
 
In this work, we compute both $\Lambda_{\rm MSR}$ using the arithmetic mean as in Allison et al. (2009a), and $\Gamma_{\rm MSR}$ using the geometric mean following Olczak et al. (2011). In each case, this is done 100 times in order to obtain the quantities $\langle l^{\rm rand}_{\rm MST} \rangle$ and $\langle \gamma^{\rm rand}_{MST} \rangle$. A value of $\Lambda_{\rm MSR} \approx 1$ (respectively $\Gamma_{\rm MSR} \approx 1$) implies that both samples of stars (i.e., the most massive and the randomly selected) are distributed in a similar manner, whereas $\Lambda_{\rm MSR} \gg 1$ (respectively $\Gamma_{\rm MSR} \gg 1$) indicates mass segregation, and $\Lambda_{\rm MSR} \ll 1$ (respectively $\Gamma_{\rm MSR} \ll 1$) points to inverse mass segregation, i.e. the massive stars are more spread outwards than the rest.

Since the vast majority of the clusters in the sample have ages that are much larger than a few million years and are therefore unaffected by extinction effects, we use the magnitudes of stars as a proxy for the mass. This also has the advantage of avoiding to introduce additional uncertainties when converting the observed luminosities into masses. Fig.~\ref{fig1} displays the dependence of $\Lambda_{\rm MSR}$ (top row) and $\Gamma_{\rm MSR}$ (bottom row) on $n_{\rm MST}$ for the different bands available in the MWSC, namely the $B$, $V$, $J$, $H$, and $K_{\rm s}$ bands for two selected clusters. The figure displays the case of a mass segregated cluster (MWSC 2202, left column) and of non-mass segregated cluster (MWSC 0141, right column). We observe that the level of mass segregation is insensitive to the choice of wavelength. For the remaining clusters in our sample (1276 in total, see \S.~\ref{appdata} below), we calculate $\Lambda_{\rm MSR}$ and $\Gamma_{\rm MSR}$ using the $J$ band observations. Although the number of stars in each cluster varies greatly in the sample (between a few tens to more than 4000), all clusters show a similar behaviour. We have verified that if a cluster shows evidence of mass segregation, this is usually seen only for $n_{\rm MST} \lesssim 20$, regardless of the total number of cluster members. Therefore, $n_{\rm MST}=10$ is a well justified choice for comparing different clusters. Hereafter, we will refer to the $\Lambda_{\rm MSR}$ and $\Gamma_{\rm MSR}$ parameters as $\Lambda^{J}_{10}$ and $\Gamma^{J}_{10}$.

\subsection{Application to the data}\label{appdata}

In order to minimise biases and selection effects, we only consider clusters closer than $2$ kpc from the Sun. With a decreasing number of stars in a cluster, the error on \Q\ increases, and the \Q\ values become less reliable (e.g., Gouliermis et al. 2012). Also the MSR analysis requires a minimum number of objects to give meaningful results. Therefore, we consider only those clusters with $40$ or more members (where $\sigma_\Q\ \lesssim 10\%$). Applying the restrictions ($d < 2$ kpc; a minimum number of stars of 40 with $P > 61$\% in a cluster) leaves 1276 clusters that are used in this study.

\section{Results and Discussion}\label{secresults}

\begin{figure*}
\begin{center}
\epsfig{figure=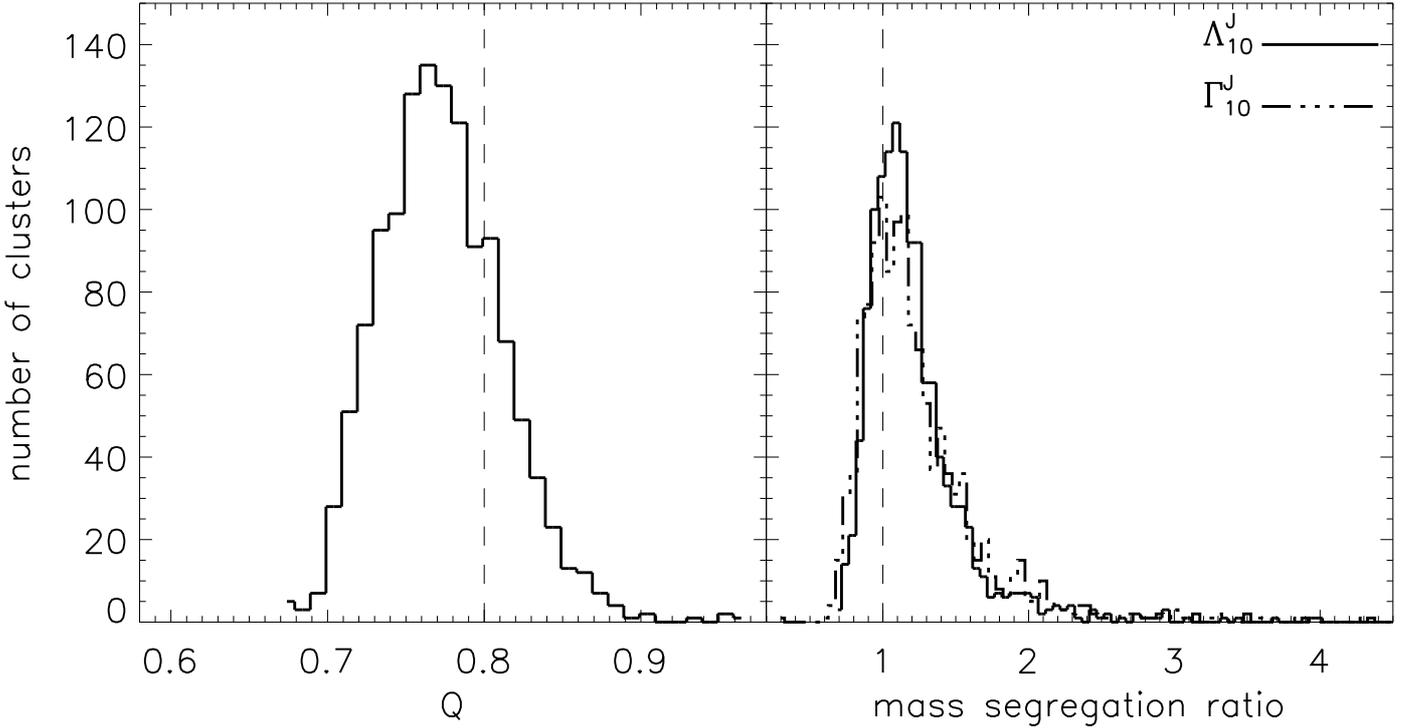,width=\textwidth}
\end{center}
\vspace{0.5cm}
\caption{Distribution of the structure parameter \Q\ (left panel) and of the mass segregation ratios $\Lambda^{J}_{10}$ and $\Gamma^{J}_{10}$ (right panel). The vertical dotted lines show the division between hierarchical and centrally concentrated clusters at $\Q\ = 0.8$ and the division between non-mass-segregated and mass-segregated clusters at $\Lambda^{J}_{10}, \Gamma^{J}_{10}=1$, respectively.}
\label{fig2}
\end{figure*}

We calculated \Q\ and the MSR ($\Lambda^{J}_{10}$ and $\Gamma^{J}_{10}$) for all of the 1276 clusters in the sample. Figure~\ref{fig2} displays the distribution of the \Q\ parameter for the entire sample (left panel). The values of \Q\ lie in the range $0.67 <$ \Q\ $< 0.97$ with an arithmetic mean value and standard deviation of  $\bar{\Q}= 0.78 \pm 0.04$. Only 344 clusters (26.95\%) possess a value of \Q\ $ > 0.8$, i.e. are centrally concentrated. The majority of clusters (72.1\%) lie in the range $0.7 <$ \Q\ $< 0.8$, showing neither central concentration nor significant substructure. This is also the range expected for a random distribution of stars. Figure~\ref{fig2} (right panel) also displays the distribution of the MSRs $\Lambda^J_{\rm 10}$ and $\Gamma^{J}_{10}$ for the sample. The distributions of $\Lambda^{J}_{10}$ and $\Gamma^{J}_{10}$ are nearly identical with the distribution of $\Gamma^{J}_{10}$ being slightly broader. Given this result, we use $\Lambda^{J}_{10}$ as a description of the MSR in the remaining sections of the paper. The values of  $\Lambda^{J}_{10}$ lie in the range $0.69 < \Lambda^{J}_{\rm 10} < 4.65$ with an arithmetic mean value and standard deviation of $\bar{\Lambda^J_{\rm 10}} = 1.23 \pm 0.37$. Only 180 clusters (14.1 \%) have values of $\Lambda^J_{10} > 1.5$ and can be considered as being significantly mass segregated. Tab.~\ref{table1} lists the parameters of the selected 1276 clusters along with their respective values of \Q\ , $\Lambda^{J}_{10}$, and $\Gamma^{J}_{10}$ derived in this work.

\begin{table*}
	\centering
	\caption{Parameters of the selected clusters in from the MWSC catalogue (1) ID in the MWSC catalogue (2) cluster name (3) R. A. (2000) (4) declination (2000) (5) Galactic Longitude (6) Galactic latitude (7) distance from the Sun (8) {\rm log}(age) (9) core radius (in pc) (10) tidal radius (in pc) (11) number of stars (12) distance to the galactic centre (13) distance to the galactic plane (14) $Q$ parameter (15) $\Lambda^{J}_{10}$ (Allison et al. 2009a) (16) $\Gamma^{J}_{10}$ (Olczak et al. 2011). The complete list for the 1276 clusters is available in the online version of the paper.}
	\begin{tabular}{llcccccccccccccc} 
		\hline
	          ID &  name &  $\alpha$ & $\delta$ & $l$ & $b$ & $D$ & ${\rm log}$ ($\tau_{cl}$) & $r_{c}$ & $r_{t}$ & $N_{*}$ & $d_{GC}$ & $z$ & $Q$ & $\Lambda^{J}_{10}$ & $\Gamma^{J}_{10}$ \\
	               &             &   [deg]        & [deg]      &  [deg] & [deg] & [pc] & [yr]                                 &  [pc]      & [pc]        &                & [pc]             & [pc]   &          &                                      & \\ 
		\hline		
		2   &   NGC 7801  &  0.082  &   50.727 &  114.717  &  -11.331 &  1953  & 9.255  &  0.61 &  9.93 &   64 &  9301 &  -384  &  0.747 &  1.360 &  1.517 \\ 
                  5    &  Berkeley 59 &  0.559  &  67.425  &  118.219  &  5.001    & 1000   &  6.100 &  0.55 &  6.51 &   88 &  8971 &   87    &  0.834 &  2.167 &   2.201\\
                  6   &   Cep OB4     &  0.735  &  67.500  & 118.299   &  5.062    &   850   &  6.100 &  9.540 &  13.56 &   653 &   8901 &    75 &   0.758 &  1.308 &  1.835\\                  
		\hline 
	\end{tabular}
	\label{table1}
\end{table*}

\subsection{Correlation of cluster structure and mass segregation with cluster age}\label{correlate}

\begin{figure}
\epsfig{figure=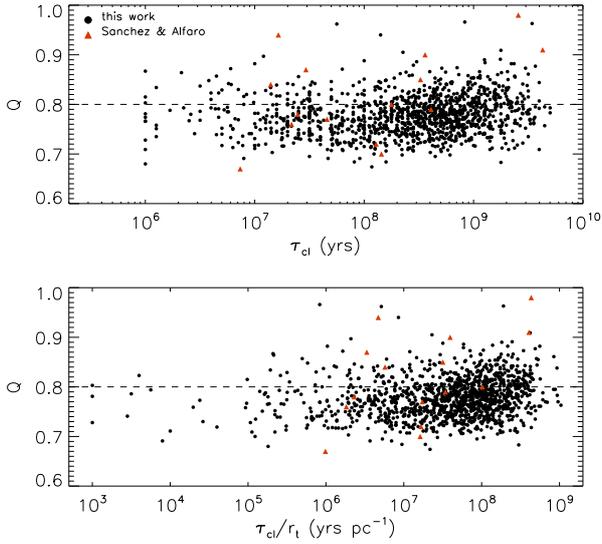,width=\columnwidth}
\caption{The top panel displays \Q\ as a function of cluster age, $\tau_{cl}$. The horizontal dotted line at \Q\ $= 0.8$ indicates the division between hierarchical and centrally concentrated clusters. The lower panel displays \Q\ plotted as a function of the ratio of the cluster age to its tidal radius. The purple symbols show the values found by S\'{a}nchez \& Alfaro (2009).}
\label{fig3}
\end{figure}

\begin{figure}
\epsfig{figure=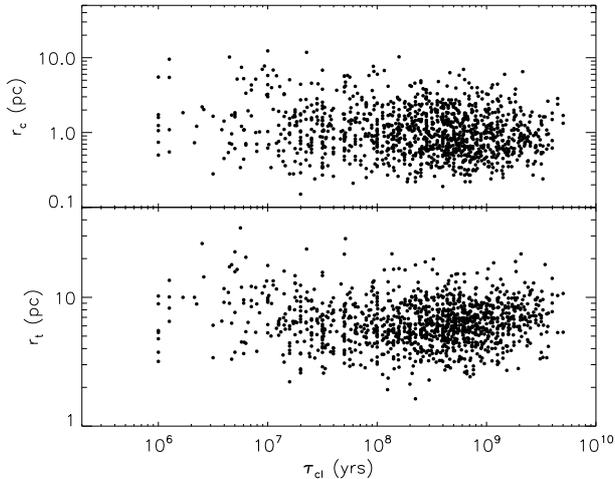,width=\columnwidth}
\vspace{0.5cm}
\caption{Core radius (top panel) and tidal radius (bottom panel) as a function of cluster age.}
\label{fig4}
\end{figure}

\begin{figure}
\epsfig{figure=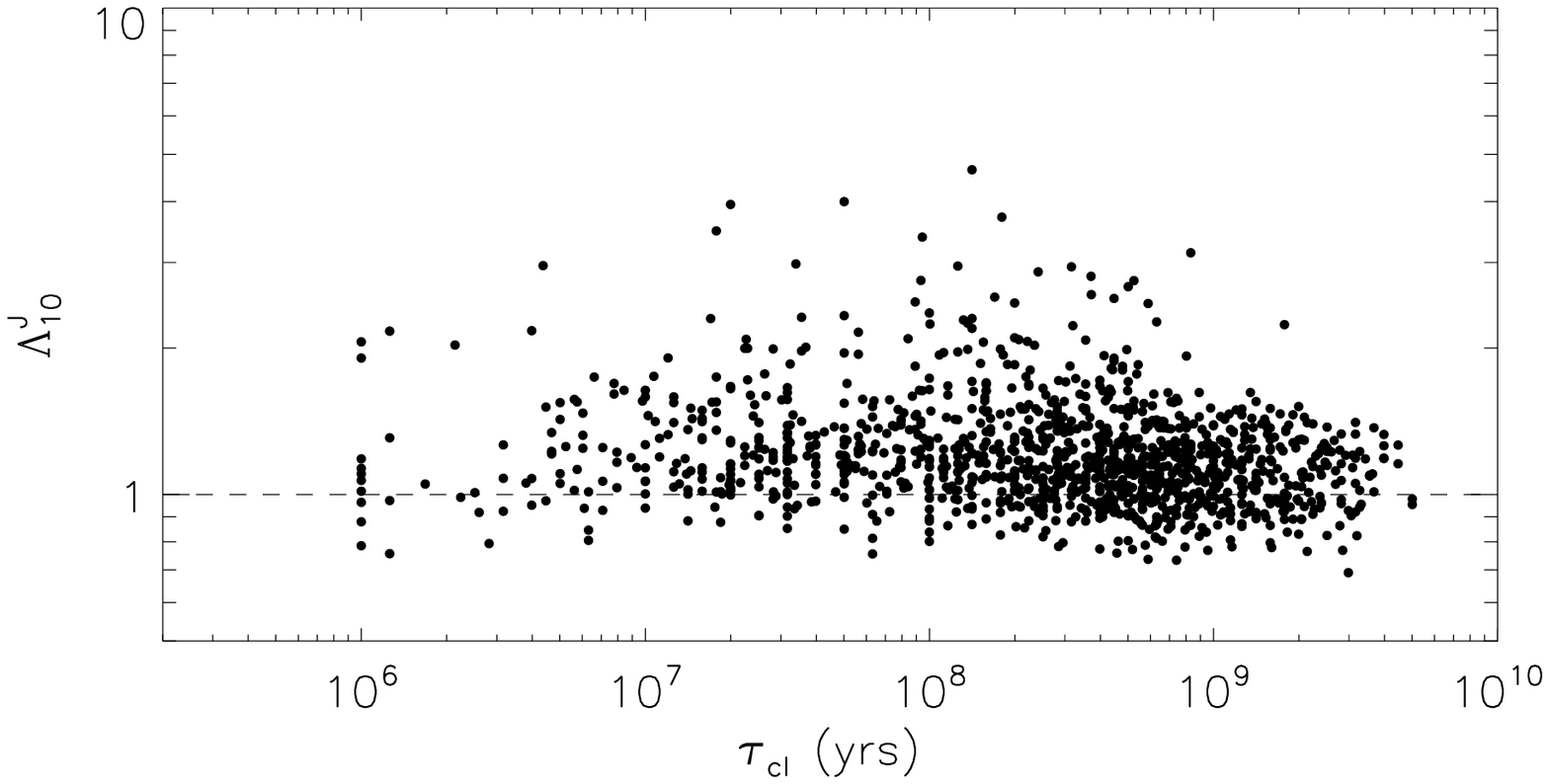,width=\columnwidth}
\caption{Mass segregation ratio $\Lambda^J_{\rm 10}$ as a function of cluster age.}
\label{fig5}
\end{figure}

The structure of clusters changes, from the onset of their formation and thorough their subsequent dynamical evolution (e.g., Schmeja \& Klessen 2006; Schmeja et al. 2008a; Parker 2014). In gravitationally bound clusters, self-gravity leads to a centrally condensed configuration, while gravitationally unbound clusters will approach nearly homogeneous distributions (with $\Q\ \approx 0.8$). It may take several crossing times to reach an equilibrium state (Goodwin \& Whitworth 2004). Simulations (Schmeja \& Klessen 2006; Moeckel \& Bate 2010; Parker \& Alves de Oliveira 2017) and observations (Schmeja et al. 2008b) indicate an increase of \Q\ during the first few Myr of a forming cluster. In their simulations, Parker et al. (2014) found this behaviour of \Q\ only for subvirial star-forming-regions, while in supervirial regions \Q\ stays at a constant low level. Figure~\ref{fig3} (top panel) displays the \Q\ values of our sample as a function of cluster age, $\tau_{cl}$. No correlation between \Q\ and $\tau_{cl}$ is observed. S\'{a}nchez \& Alfaro (2009) studied a small sample of 16 open clusters spanning a wide range of ages ($6.9 < \log(\tau_{cl}) < 9.6$), and determined their \Q\ values (red triangles in Fig.~\ref{fig3}). The latter authors argued for a weak correlation of \Q\ with age, but their conclusion is not substantiated by our findings. They also argued that a correlation exists between \Q\ and the ratio of the cluster age divided by the tidal radius which is proportional to the age of the cluster expressed in units of the crossing time. They find the relation \Q\ $=(0.07 \pm 0.03) \log (\tau_{cl}/r_{t}) + (0.35 \pm 0.21)$ where $r_{t}$ is the tidal radius. We seek the same correlations between \Q\ and $\tau_{cl}/r_{t}$ in our sample. The result is displayed in Fig.~\ref{fig3} (bottom panel) along with the data points of S\'{a}nchez \& Alfaro (2009). Our data do not suggest the existence of a correlation between \Q\ and $(\tau_{cl}/r_{t})$. In fact, most of the \Q\ values for the clusters of the MWSC lie far below the correlation suggested by S\'{a}nchez \& Alfaro (2009). We attribute this discrepancy to the different samples and to the small number of clusters studied by S\'{a}nchez \& Alfaro (2009). We also note a large difference in the \Q\ parameter (up to $\Delta$\Q\ $ \approx 0.25$) for a few clusters (for example for the cluster MWSC 3008, we find a value of  \Q\ $=0.77$, whereas S\'{a}nchez \& Alfaro found \Q\ $=1.02$. A possible interpretation of the absence of a correlation between \Q\ and cluster age (or between \Q\ and $\tau_{cl}/r_{t}$) implies that even though if it is likely that \Q\ increases with time for individual clusters, at least in the early formation period, clusters start from having different \Q\ values and follow a distinct individual evolution, such that a general correlation for all clusters is not to be expected. We should also point out that if we were missing stars that are located in the outskirts of the clusters (i.e., outliers), this would have the effect of artificially decreasing the \Q\ parameter. The effect of missing outliers is difficult to quantify, because obviously this effect may depend on their numbers and spatial distributions. An example of this effect for the young clusters IC 348 and NGC 1333 is demonstrated in Parker \& Alves de Oliveira (2017). The \Q\ parameter for these two clusters decreases by $0.1-0.15$ when the outer regions are omitted in the calculations. 

Tadross (2014) found a weak correlation between the age of a cluster and its diameter and Schilbach et al. (2006) found a dependence of cluster radius on age. However, they attributed this dependence to the effects of mass segregation which are ubiquitous for clusters older than $30$ Myrs in their sample. Figure~\ref{fig4} displays the core radius, $r_{c}$ (top panel) and the tidal radius, $r_{t}$, (bottom panel) plotted as a function of the cluster age, $\tau_{cl}$. There are no visible correlations between $r_{c}$ or $r_{t}$ with the cluster age. 

A correlation of mass segregation levels with age may be expected from the dynamical evolution of clusters. However, this effect may be overshadowed by the existence of different levels of primordial mass segregation in the clusters (Dib et al. 2007a; Dib et al. 2010). From a dynamical point of view, mass segregation can occur on short timescales of a few Myr, leading to a rapid rise of $\Lambda_{\rm MSR}$ (e.g., Allison et al. 2009b). A similar trend for $\Lambda_{\rm MSR}$ has been observed by Parker et al. (2014). In their simulations, supervirial regions show no sign of mass segregation, i.e., $\Lambda_{\rm MSR}$ stays at unity for the entire time of the simulation. On the other hand, subvirial regions show a wide variety in the evolution of $\Lambda_{\rm MSR}$. Usually, $\Lambda_{\rm MSR}$ increases over the first few Myr due to dynamical mass segregation with values up to $\Lambda_{\rm MSR} \approx 10$, after which, $\Lambda_{\rm MSR}$ can evolve in many different ways. In some models, $\Lambda_{\rm MSR}$ remains at high values, while in others it drops again to $\Lambda_{\rm MSR} \approx 1$. On the observational side, Bukowiecki et al. (2012) used a sample of 599 open clusters selected from the {\it Two Micron All Sky Survey} (2MASS) and argued that there is a tendency of mass segregation to increase with age. Fig.~\ref{fig5} displays the values of $\Lambda_{10}^{J}$ plotted as a function of cluster age for the sample of 1276 clusters used in this study. No correlation is visible between $\Lambda_{10}^{J}$ and age.

\subsection{Correlation of cluster structure and mass segregation with Galactic position}\label{correlpos}

\begin{figure}
\epsfig{figure=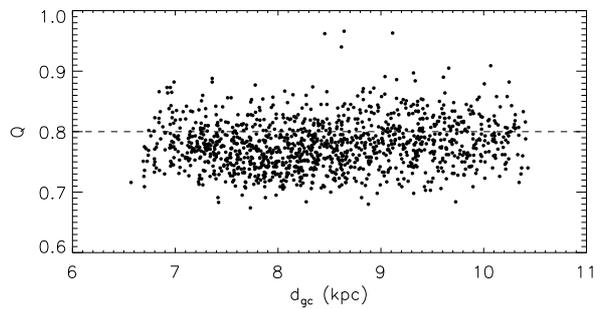,width=\columnwidth}
\caption{\Q\ as a function of Galactocentric distance. The Sun is assumed to lie at $d_{\rm GC_{\odot}} = 8.5$~kpc.}
\label{fig6}
\end{figure}

\begin{figure}
\epsfig{figure=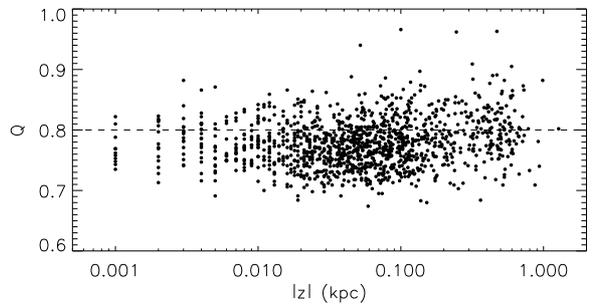,width=\columnwidth}
\caption{\Q\ as a function of the distance $|z|$ from the Galactic plane.}
\label{fig7}
\end{figure}

\begin{figure}
\epsfig{figure=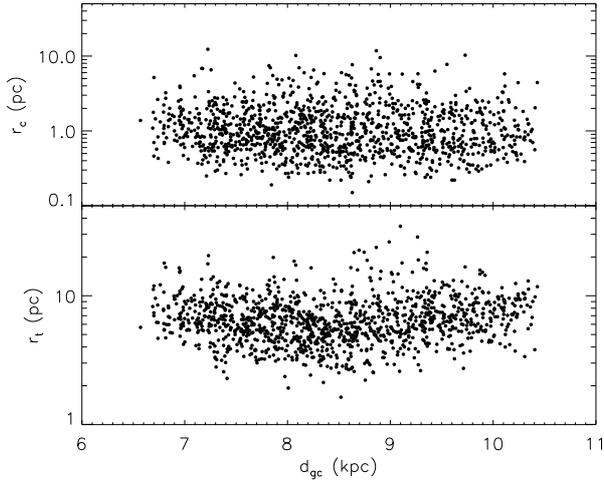,width=\columnwidth}
\vspace{0.5cm}
\caption{Core radius (top panel) and tidal radius (bottom panel) as a function of Galactocentric distance $d_{\rm GC}$.}
\label{fig8}
\end{figure}

\begin{figure}
\epsfig{figure=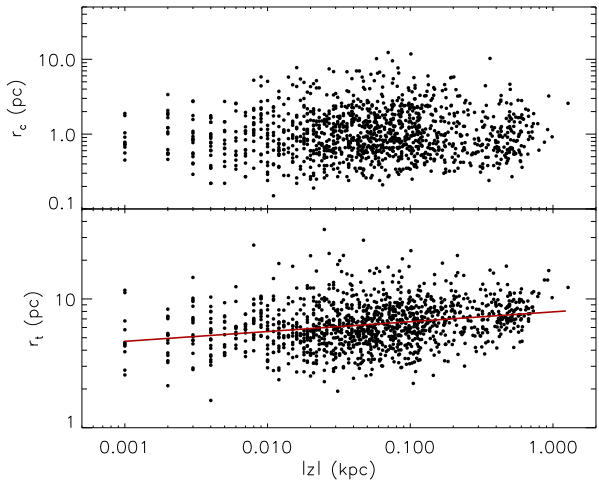,width=\columnwidth}
\vspace{0.5cm}
\caption{Core radius (top panel) and tidal radius (bottom panel) as a function of $|z|$. The red line shows a linear fit to the data. The parameters of the fit are reported in the text.}
\label{fig9}
\end{figure}

\begin{figure}
\epsfig{figure=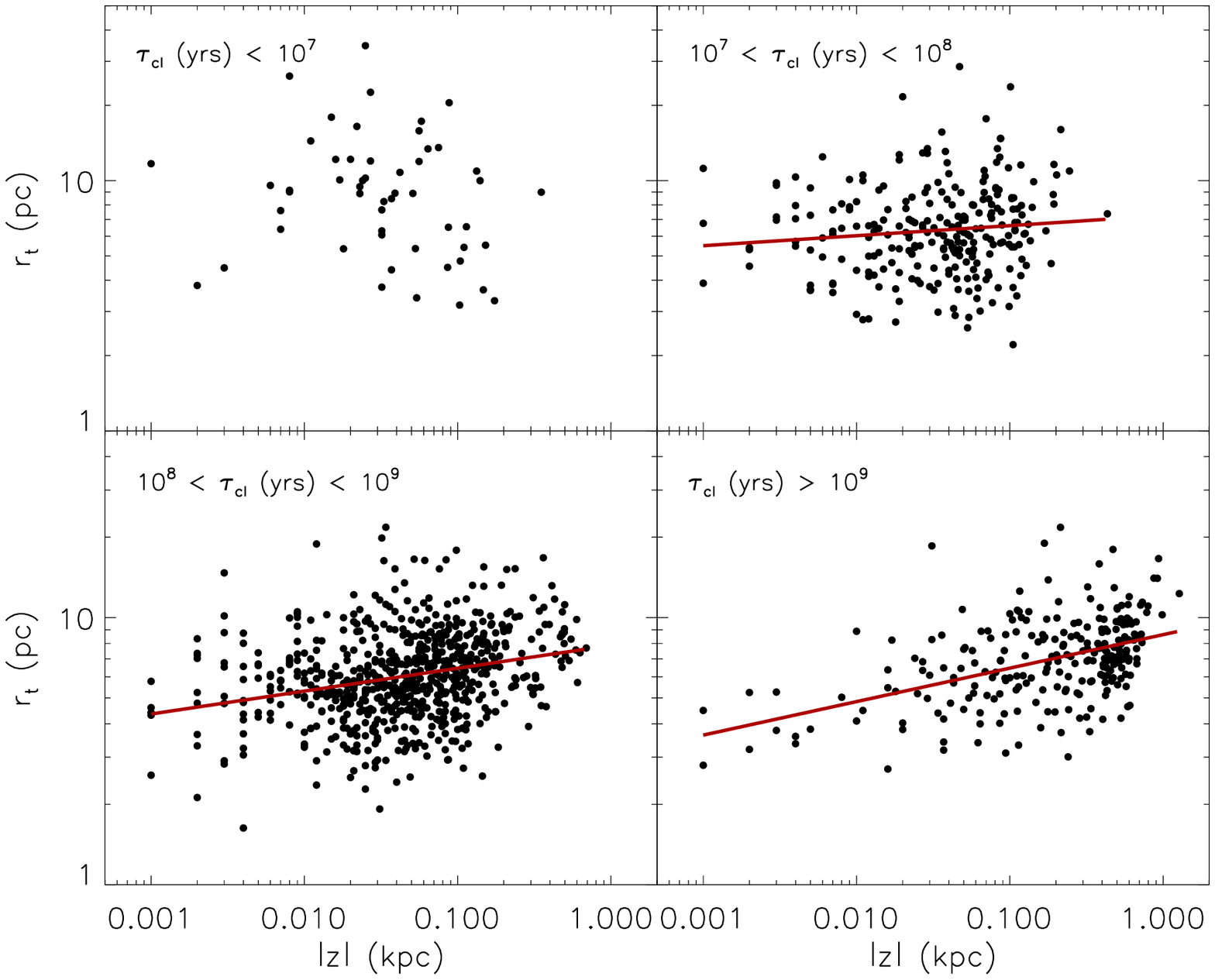,width=\columnwidth}
\vspace{0.5cm}
\caption{$|z|$ versus tidal radius in four age bins. The red lines show linear fits to the data. The parameters of the fits are reported in the text.}
\label{fig10}
\end{figure}

As the environment may have an influence on the structure of the clusters, we investigate the existence of potential correlations of the clusters parameters with their position in the Galaxy, characterised by the Galactocentric distance $d_{\rm GC}$, the distance from the Galactic plane $|z|$, and the location of the clusters in or outside of the spiral arms. Tadross (2014) found a slight correlation between Galactocentric radius and $|z|$ with the size of the clusters. Froebrich et al. (2010) noted that more extended clusters are found more often at large Galactocentric distances as well as at larger $|z|$. Schilbach et al. (2006) found a systematic increase of cluster size with $|z|$, which becomes significant for clusters older than $\log(\tau_{cl}) = 8.35$.

In our sample, \Q\ does not show any correlation with $d_{\rm GC}$ or $|z|$ (see Fig.~\ref{fig6} and Fig.~\ref{fig7}). Neither do we find a correlation of the core or tidal radii with $d_{\rm GC}$ (Fig.~\ref{fig8}). While no correlation is observed between $r_{c}$ and $|z|$ (Fig.~\ref{fig9}, top panel), we do however, find a correlation of $r_{t}$ with $|z|$ (Fig.~\ref{fig9}, bottom panel). The correlation is given by ${\rm log} (r_{t}) = 0.076(\pm 0.008)~{\rm log}|z| + 0.89(\pm 0.01)$ (with a Pearson correlation coefficient of $\approx 0.25$). This is in agreement with the findings of Schilbach et al. (2006), Froebrich et al. (2010), Bukowiecki et al. (2011), and Tadross (2014). We analysed the same relation for different age bins (Fig.~\ref{fig10}). While there is no obvious correlation in the age bin $\log(\tau_{cl}) [{\rm yr}] < 7$, we find a correlation between $|z|$ and the tidal radius for ages  $7 < \log(\tau_{cl}) [{\rm yr}] < 8$ (${\rm log} (r_{t}) = 0.040(\pm0.023)~{\rm log}|z| + 0.86(\pm0.03)$), $8 < \log(\tau_{cl}) [{\rm yr}] < 9$ (${\rm log} (r_{t}) = 0.085(\pm0.011)~{\rm log} |z| + 0.89\pm(0.01)$), and $\log(\tau_{cl}) [{\rm yr}] \ge 9$ (${\rm log} (r_{t}) = 0.12(\pm0.015)~{\rm log}|z| + 0.93(\pm0.01)$). The Pearson correlation coefficients are $0.10$, $0.27$, and $0.46$ for the age bins $[10^{7}-10^{8}]$ yrs, $[10^{8},10^{9}]$ yrs, and $> 10^{9}$ yrs, respectively, indicating an increase in the correlation between $r_{t}$ and $|z|$ with increasing age.   

\begin{figure}
\epsfig{figure=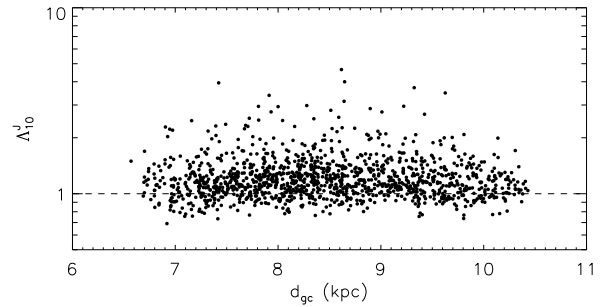,width=\columnwidth}
\caption{MSR parameter ($\Lambda^{J}_{10}$) versus Galactocentric distance. The Sun is assumed to lie at $d_{\rm GC_{\odot}} = 8.5$~kpc).}
\label{fig11}
\end{figure}

\begin{figure}
\epsfig{figure=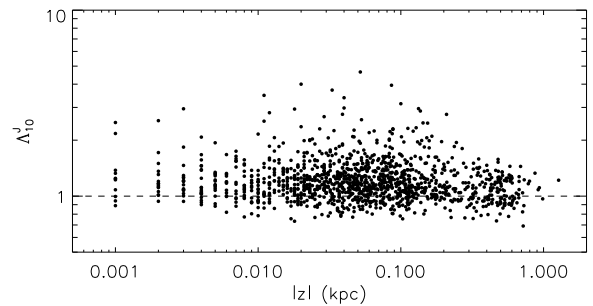,width=\columnwidth}
\caption{MSR parameter ($\Lambda_{10}^{J}$) versus distance $|z|$ from the Galactic plane.}
\label{fig12}
\end{figure}

\begin{figure}
\epsfig{figure=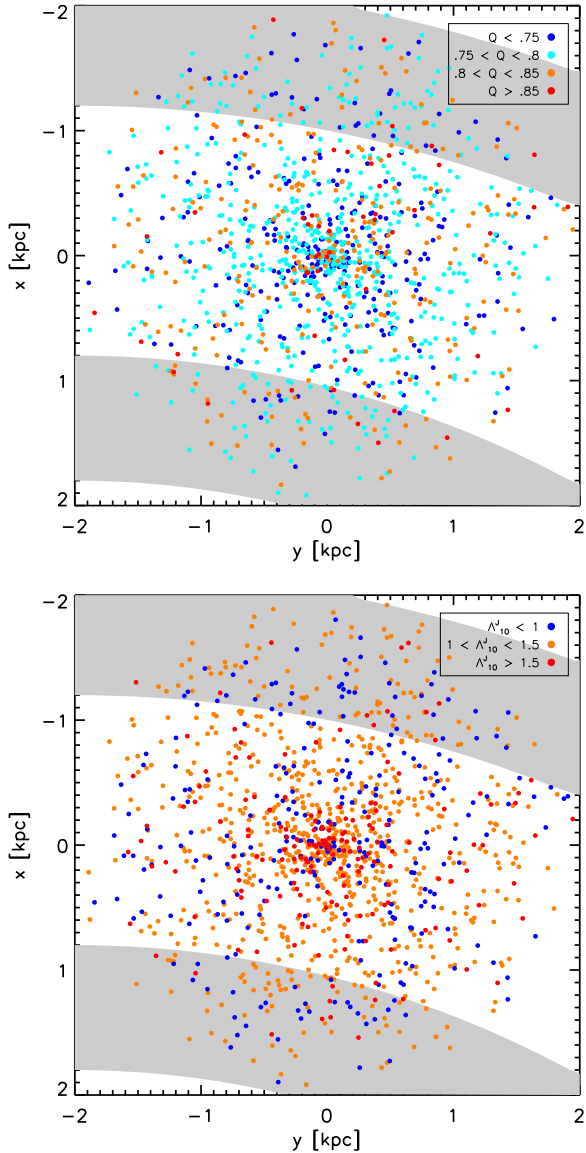,width=\columnwidth}
\caption{Spatial distribution of the clusters centred around the Galactic position of the Sun. The grey areas are the inner (Sagittarius) and outer (Perseus) spiral arms around the Sun. The clusters are coded by the value of their structure parameter \Q\ (top panel), and their mass segregation ratio $\Lambda^{J}_{10}$ (bottom panel). There are no obvious correlations between the \Q\ or $\Lambda^{J}_{10}$ values of the clusters with their position in the arms or in the interarm regions.}
\label{fig13}
\end{figure}

$\Lambda_{10}^{J}$ shows no correlation with $d_{\rm GC}$ (Fig.~\ref{fig11}). Higher values of $\Lambda_{10}^{J}$ ($\gtrsim 2$) are only found for cluster at smaller $|z|$ ($|z| \lesssim 250$ pc (Fig.~\ref{fig12}). This is likely to be only a statistical effect, since there are many more clusters close to the Galactic plane than at high $|z|$. However, the mean value of $\Lambda_{10}^{J}$ for consecutive bins containing each 50 clusters does not change significantly with $|z|$ (not shown). We also test whether the cluster parameters show any dependence with respect to their location inside/outside of the Galactic spiral arms (Fig.~\ref{fig13}). The positions of the spiral arms (Perseus and Sagittarius arms) are taken from Vall\'{e}e (2014). The average \Q\ parameter is exactly the same, $\bar{\Q}=0.78 \pm 0.04$, for the clusters inside and outside the arms. The $\Lambda_{10}^{J}$ values are on average slightly higher outside the spiral arms ($\bar{\Lambda_{10}^{J}}=1.25 \pm 0.38$) than inside ($\bar{\Lambda_{10}^{J}}=1.18 \pm 0.33$), but these values are compatible within the $1\sigma$ uncertainty. Of the clusters outside the arms, $\approx 16\%$ of them show significant mass segregation ($\Lambda_{10}^{J} > 1.5$), while this is only the case for $\approx 7\%$ of all the clusters in the spiral arms. The average tidal radii also do not show a significant difference inside ($\bar{r_{t}}=7.36 \pm 2.81$ pc) and outside ($\bar{r_{t}}=6.73 \pm 3.23$ pc) the spiral arms.

\subsection{Implications for star formation in the local volume}\label{implicationsf}

\begin{figure*}
\epsfig{figure=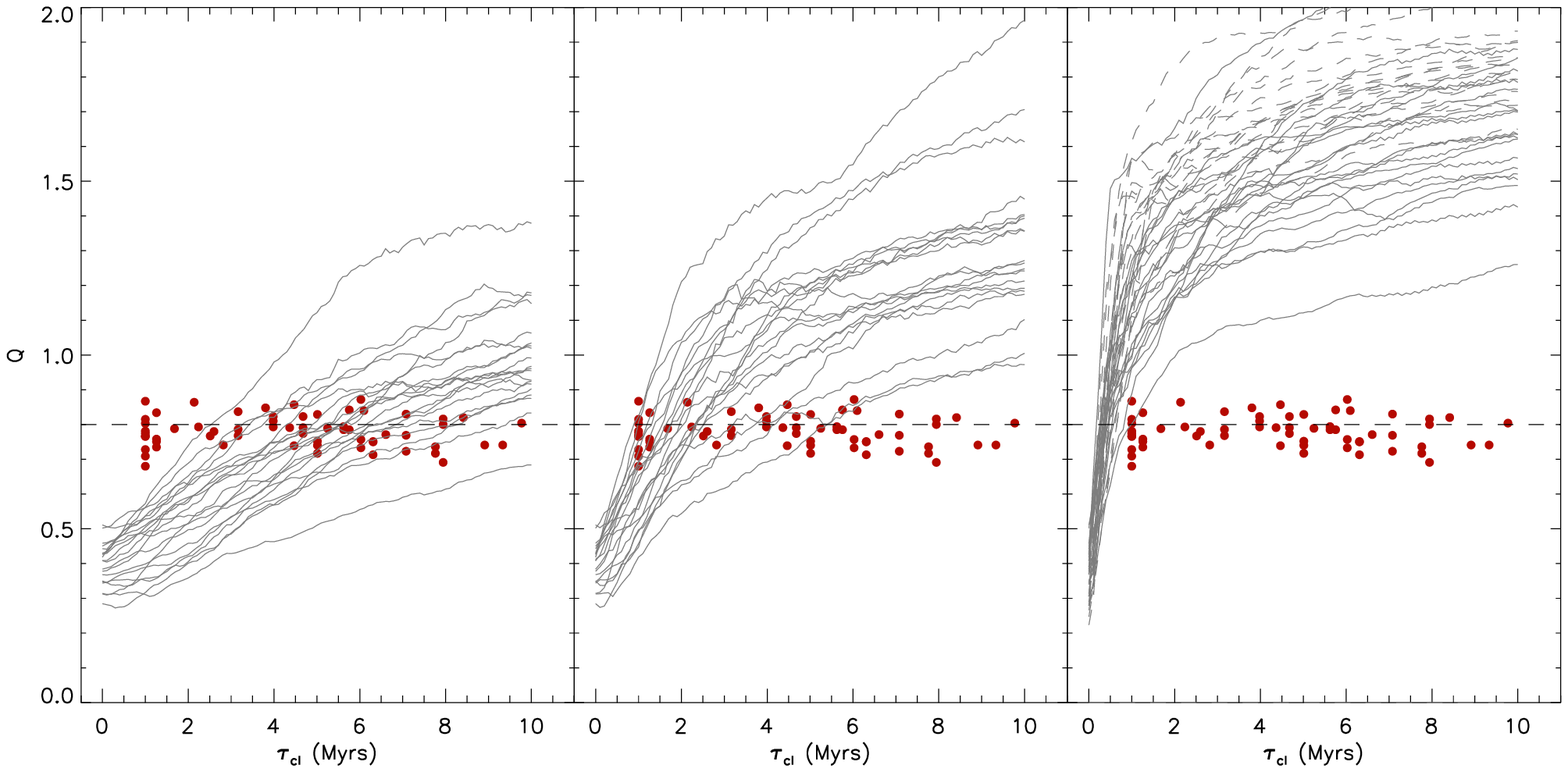,width=\textwidth}
\caption{Comparison of the \Q\ parameter for the young clusters in the MWSC to the evolution of \Q\ in N-body models of stellar clusters. The N-body models includes ones for clusters with $N_{*}=425$ in which the initial local volume densities in the clusters falls in the range 10-60 M$_{\odot}$ pc$^{-3}$ (full line, left panel), 100-500 M$_{\odot}$ pc$^{-3}$ (full line, middle panel), and 5000-10000 M$_{\odot}$ pc$^{-3}$ (full line, right panel) and with $N_{*}=1500$ with initial local volume densities that are in the range $2\times10^{3}-10^{4}$ M$_{\odot}$ pc$^{-3}$ (dashed line, right panel).}
\label{fig14}
\end{figure*}

It is possible to gain insight into the physical conditions prevalent at the time the young clusters in the MWSC formed by comparing their structure and mass segregation levels with those derived from numerical simulations. Magnetohydrodynamical simulations of star cluster formation have the advantage of taking into account the effects of the gas on the structure of the nascent clusters and can follow the evolution of the cluster properties during its build up. However, star cluster formation simulations are computationally expensive and can therefore sample only a limited subset of the the initial conditions of the parent protocluster clumps (e.g., Dib et al. 2007b; Dib et al. 2008b; Dib et al. 2010b; Padoan et al. 2014). An appealing alternative is to use N-body simulations which can follow the evolution of a cluster over an extended period of time. These simulations can start either from a gas free cluster and can use as initial conditions of the stars positions and kinematics the input of star formation models or be constructed with more controlled and idealised initial conditions. Parker (2014), Parker et al. (2014, 2015) and Parker \& Alves de Oliveira (2017) presented a number of such simulations. Parker et al. (2014) simulated the dynamical evolution of initially hierarchically structured clusters over the first 10 Myr starting from different initial conditions, and followed the evolution of \Q\ and $\Lambda_{\rm MSR}$. These N-body simulations explored the effect of starting from subvirial or supervirial conditions and the effect of a different initial fractal dimension of the clusters. Parker et al. (2014) found that in subvirial regions ($\alpha_{\rm vir}$ in the range  $[0.3-0.5]$), substructure is erased rapidly and \Q\ rises to values $> 1$ within 1 Myr. On the other hand, in supervirial regions ($\alpha_{\rm vir} = 1.5$), substructure is preserved and a constant low \Q\ characteristic of cluster with substructure in maintained. Cases of clusters that have $\alpha_{\rm vir} \approx 0.5-1.5$ are not yet fully explored and could display an intermediate behaviour, i.e., a moderate rise in the value of \Q\ followed by a saturation at that level.  

Parker (2014) explored the effect of changing the initial cluster density\footnote{This is a working assumptions as stars will not form simultaneously. Instead this concept of initial stellar density could be understood as an initial peak stellar density.} ($\rho_{*}$) of the star-forming region on the evolution of the \Q\ parameter (figure 3 in his paper). High-density regions ($\rho_{*} \approx 10^4$ M$_{\odot}$ pc$^{-3}$) lose substructure within 1 Myr and reach values $\Q > 1$ after 10 Myr (i.e., at the end of the simulations), medium-density regions ($\rho_{*} \approx 10^2$ M$_{\odot}$ pc$^{-3}$) lose substructure within $3-5$ Myr and end up with $0.7 \lesssim \Q \lesssim 1.2$, whereas low-density regions ($\rho_{*} \approx 10$ M$_{\odot}$ pc$^{-3}$) retain substructure for the entire time and stay at values $0.4 \lesssim \Q \lesssim 0.7$.

In this work we compare the \Q\ values derived for the population of young clusters in the MWSC to a set of N-Body models presented in Parker \& Alves de Oliveira (2017). These models follow the time evolution of the \Q\ parameter in clusters with $N_{*}=425$ stars\footnote{As the \Q\ parameter displays a slight dependence on the number of stars (e.g., Lomax et al. 2011), ideally, each cluster in the observational sample should be compared to simulations that are performed using the same number of stars. However, in practice, the scatter between N-body simulations constructed with the same set of parameters but with different random seed number for the spatial and kinematic distributions of the stars is larger than the effect of the number of stars.} and which have an initial virial ratio of $\alpha_{\rm vir}=0.3$. The clusters are initially sub-structured and have a fractal dimension $D=1.6$. The models include cases with initial cluster radii of 0.5, 1.5, and 3 pc. The stellar masses of the 425 stars are drawn from the observed IMF of IC 348 (Luhman et al. 2016)\footnote{The Luhman et al. (2016) paper list the photometric data of the 425 stars in IC 348. The stellar masses have been derived in Parker \& Alves de Oliveira (2017) using the Luhman et al. (2016) data following a procedure described in detail in Section 2 of their paper.} and the initial positions of the stars within the cluster are randomly assigned and no correlation between the masses of the stars and their positions within the cluster is imposed. For the three chosen values of the cluster initial radii, this leads to local volume densities in the clusters that fall in the range $\rho_{*}\approx 10-60$ M$_{\odot}$ pc$^{-3}$ when the radius is $3$ pc, $100-500$ M$_{\odot}$ pc$^{-3}$ for a cluster radius of $1.5$ pc, and $5000-10000$ M$_{\odot}$ pc$^{-3}$ for a cluster radius of $0.5$ pc. We also include a set of simulations which have a higher number of stars $N_{*}=1500$, whose masses are randomly drawn from a Galactic field like IMF (Maschberger 2013)\footnote{The functional form for the IMF proposed by Maschberger (2013) is an order-3 Logistic function which is described by three parameters, namely the slope in the low mass regime, the slope in the intermediate to high mass regime, and a parameter that ensure the continuity across these two mass regimes.} and whose spatial positions are randomly assigned within the fractal structure. This additional set of simulations are performed with $\alpha_{\rm vir}=0.3$, $D=1.6$, and an initial cluster radius of $1$ pc, corresponding to initial local volume densities that fall in the range $\rho_{*} \approx 2\times10^{3}-10^{4}$ M$_{\odot}$ pc$^{-3}$. The simulations do not have stellar evolution switched on, nor do they feature an external Galactic tidal field. Simulations used in Parker et al. (2016) find no appreciable difference between the long-term dynamical evolution of clusters with or without stellar evolution.

In Fig.~\ref{fig14} we compare the \Q\ values in the observations and the N-body models. The \Q\ values for the young MWSC clusters which fall in the range of $0.7 \lesssim \Q \lesssim 0.9$ are better matched with the models that start with initial stellar surface densities of $\rho_{*} =10-100$ M$_{\odot}$ pc$^{-3}$ (left panel and middle panel in Fig.~\ref{fig14}). Models that possess higher stellar surface densities (right panel) fail to reproduce this set of observations. This implies that the young clusters in the MWSC were formed with roughly similar initial stellar volume densities and probably out of protoclusters clumps with similar structural and dynamical properties. Gregorio-Hetem et al. (2015) performed a similar comparison for their sample of 25 stellar associations with earlier N-Body models performed by Parker \& Dale (2013). They found that their data points are better reproduced with models that have initial volume densities of $\approx 5$ M$_{\odot}$ pc$^{-3}$. The presence of an external Galactic tidal field can expedite the dissolution of star clusters. However, the absence of a tidal field in our simulations will not affect our interpretation that the initial stellar densities were no higher than $10-100$ M$_\odot$ pc$^{-3}$ for two reasons. Firstly, initially high stellar densities ($> 1000$ M$_\odot$ pc$^{-3}$) would produce high values of \Q\ and $\Lambda_{\rm MSR}$, which we do not see in our sample of observed clusters. These high densities would lead to significant dynamical interactions during the early stages of the clusters' lives, where two body relaxation would dominate over the effects of the tidal field. Secondly, Parker et al. (2016) have recently shown that clusters do not approach energy equipartition, where the lowest mass objects would be ejected to the outskirts of the cluster. Therefore, we would not expect the influence of an external tidal field to preferentially remove low-mass objects from the cluster and bias the measurement of $\Lambda_{\rm MSR}$. We also note that even in this unlikely scenario, the \Q\ parameter would be unaffected, as it is independent of stellar mass.

\subsection{Discussion}\label{secdiscuss}

As discussed above, numerical simulations show that the \Q\ parameter can rise quickly in a star-forming cluster (e.g., Schmeja \& Klessen 2006; Moeckel \& Bate 2010, Parker et al. 2014; Parker 2014). In order to better understand the transition from the embedded phase into the gas-free phase in terms of the clusters structure, we analyse an additional sample of embedded clusters, taken from a study performed with the {\it Spitzer Space Telescope} located within 1 kpc of the Sun (Gutermuth et al. 2009). Computing the \Q\ values for this sample gives a mean value of $\Q = 0.86 \pm 0.08$ for the 20 clusters with 40 or more stars. This value is higher than the mean value for our open clusters (\Q\ $=0.78\pm0.04$). A similar value of $\Q = 0.87 \pm 0.07$ was found by (Jaehnig et al. 2015) for 22 young (ages $\approx 1-3$ Myr) clusters in Galactic star-forming regions. However, considering only the MWSC clusters with ages $< 5$ Myr (the maximum time for clusters expected to be embedded) results in a mean value of $\Q = 0.78 \pm 0.04$, exactly the same as for the entire sample. The discrepancy may be attributed to the different small samples, as well as to the problematic definition of embedded clusters (e.g., Kroupa 2011). So the notion that young clusters may have, on average, higher \Q\ values, as noted for a different sample in (Schmeja et al. 2008a), may not hold.

We check whether possible biases are induced by the cluster sample which is affected by incompleteness (Dib et al. 2017). We perform simple comparisons with the sample used in previous sections by selecting clusters that are either at distances $d < 1$~kpc (323 clusters) or $d < 0.5$~ kpc (78 clusters) from the Sun, or clusters with more than 500 members (78 clusters). The results of these tests show that this selection criteria do not change any of our results. The chosen membership probability is also not critical to the results. When varying the required membership probability between $P > 30\%$ and $P > 75\%$, \Q\ changes on average by less then 5\% and $\Lambda_{\rm 10}^J$ by 17\%, while the number of stars changes on average by $\approx 30-40\%$. Likewise using a different value of (i.e., $n_{\rm MST} = 5$ or $n_{\rm MST}=15$) or different filters for the MSR comparison (see Fig.~\ref{fig1} for two examples) does not result in a different behaviour (for more details on the effects of changing these quantities, see App.~\ref{appendixa}).

\section{Summary}\label{secconc}
We analysed 1276 Galactic open clusters with uniform astrophysical data from the Milky Way Stellar Cluster (MWSC) catalogue and computed their structure parameter \Q\ and their mass segregation ratio $\Lambda_{\rm MSR}$. Our main findings can be summarised as follows:
\begin{enumerate}
\item Most clusters possess values of the \Q\ parameter that fall in the range $0.7 < \Q < 0.8$, indicating neither central concentration nor significant substructure. Only $\approx 27\%$ can be considered centrally concentrated ($\Q > 0.8$).
\item Most clusters show mass segregation values around $\Lambda_{\rm MSR} \approx 1$, indicating a similar distribution of massive and low-mass stars. The distribution function of $\Lambda_{\rm MSR}$ is positively skewed and $\approx 14\%$ of the clusters show signs of significant mass segregation ($\Lambda_{\rm MSR} > 1.5$).
\item No correlation is found between \Q, $\Lambda_{\rm MSR}$, or the cluster radius with the cluster age. Some of the correlations claimed by other authors using much smaller cluster samples could not be confirmed.
\item No significant correlation is found between \Q, $\Lambda_{\rm MSR}$, or cluster radius with the cluster position in the Galaxy i.e., the distance from the Galactic centre $d_{\rm GC}$, the distance from the Galactic plane $|z|$, or the position in the arm/interam regions. There is a tendency for clusters at larger distances from the Galactic plane (i.e., large $|z|$) to have larger tidal radii, which holds in particular for older clusters.
\item Embedded and open clusters show on average the same \Q\ values.
\item by comparing the observed \Q\ values of the young clusters in the MWSC to a suite of N-body numerical simulations of the early evolution of stellar clusters suggests that the clusters found in the MWSC catalogue were formed from sub-virial/virial conditions and with mean local volume densities of $\rho_{*}\approx 10-100$ M$_{\odot}$ pc$^{-3}$.

\end{enumerate}

\section*{acknowledgements}

We are very grateful to the anonymous referee for constructive comments that help clarify a number of points in the paper. SD was supported by a Marie-Curie Intra European Fellowship under the European Community's Seventh Framework Program FP7/2007-2013 grant agreement no. 627008 in the early phase of this work. SD acknowledges the hospitality of the University of Western Ontario where parts of this work have been completed. SS was supported by Sonderforschungsbereich SFB 881 "The Milky Way System'' (subproject B5, funding period 2011-2014) of the German Research Foundation (DFG) during part of this work. RJP acknowledges support from the Royal Society in the form of a Dorothy Hodgkin Fellowship. This research has made use of NASA's Astrophysics Data System Bibliographic Services.

{}

\appendix 

\section{Sensitivity of the results to the choice of $\lowercase{n}_{\rm MST}$ and cluster membership probability}\label{appendixa}

\begin{figure}
\begin{center}
\epsfig{figure=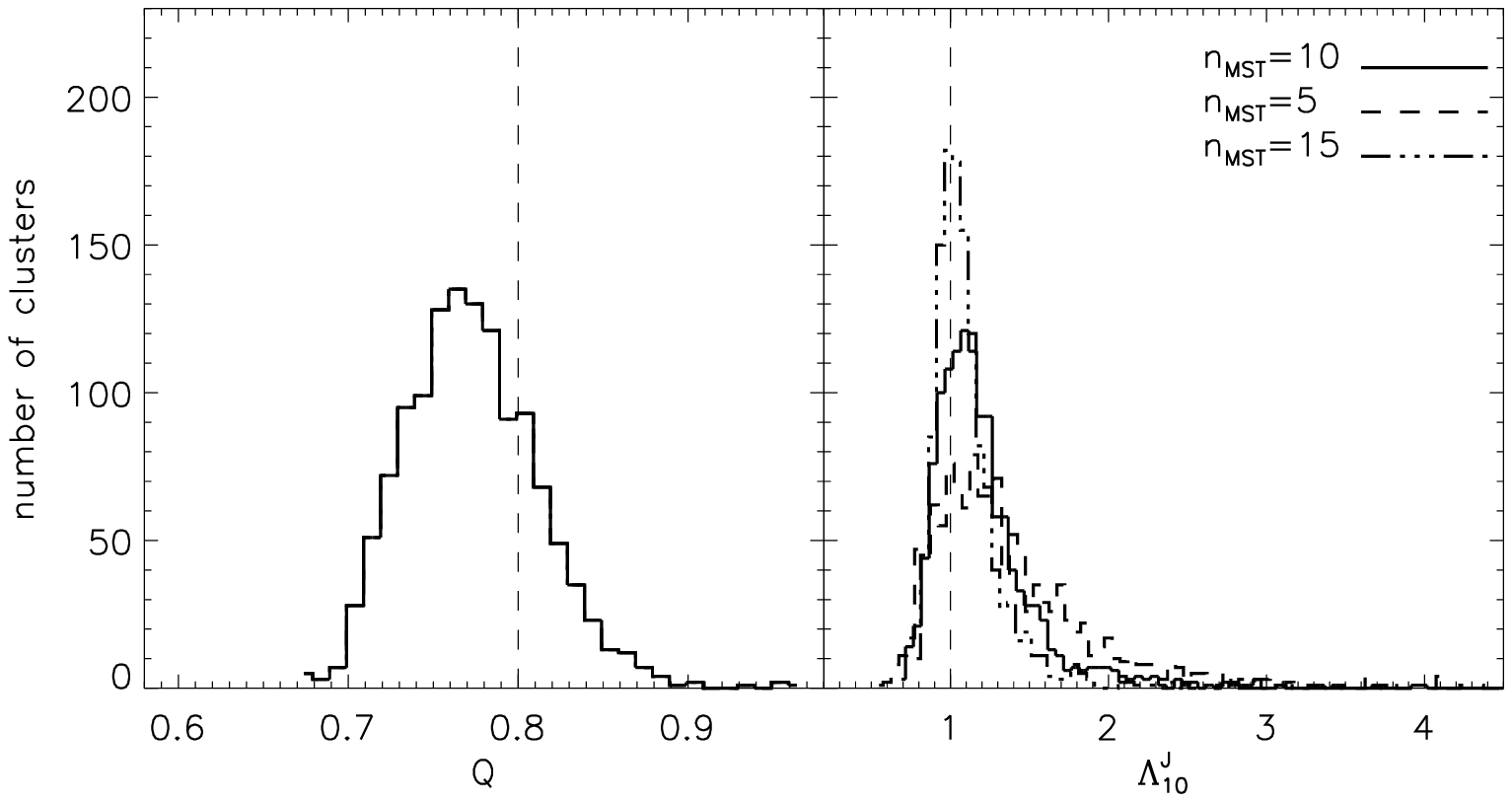,width=\columnwidth}
\end{center}
\vspace{0.5cm}
\caption{The dependence of the distribution of the structure parameter \Q\ (left panel) and of the mass segregation ratios $\Lambda^{J}_{10}$ (right panel) on the choice of $n_{\rm MST}$. The cases shown are for $n_{\rm MST}=5$, $n_{\rm MST}=10$ (fiducial case), and $n_{\rm MST}=15$. The vertical dotted lines show the division between hierarchical and centrally concentrated clusters at $\Q\ = 0.8$ and the division between non-mass-segregated and mass-segregated clusters at $\Lambda^{J}_{10}=1$, respectively.}
\label{fig1app}
\end{figure}

\begin{figure}
\begin{center}
\epsfig{figure=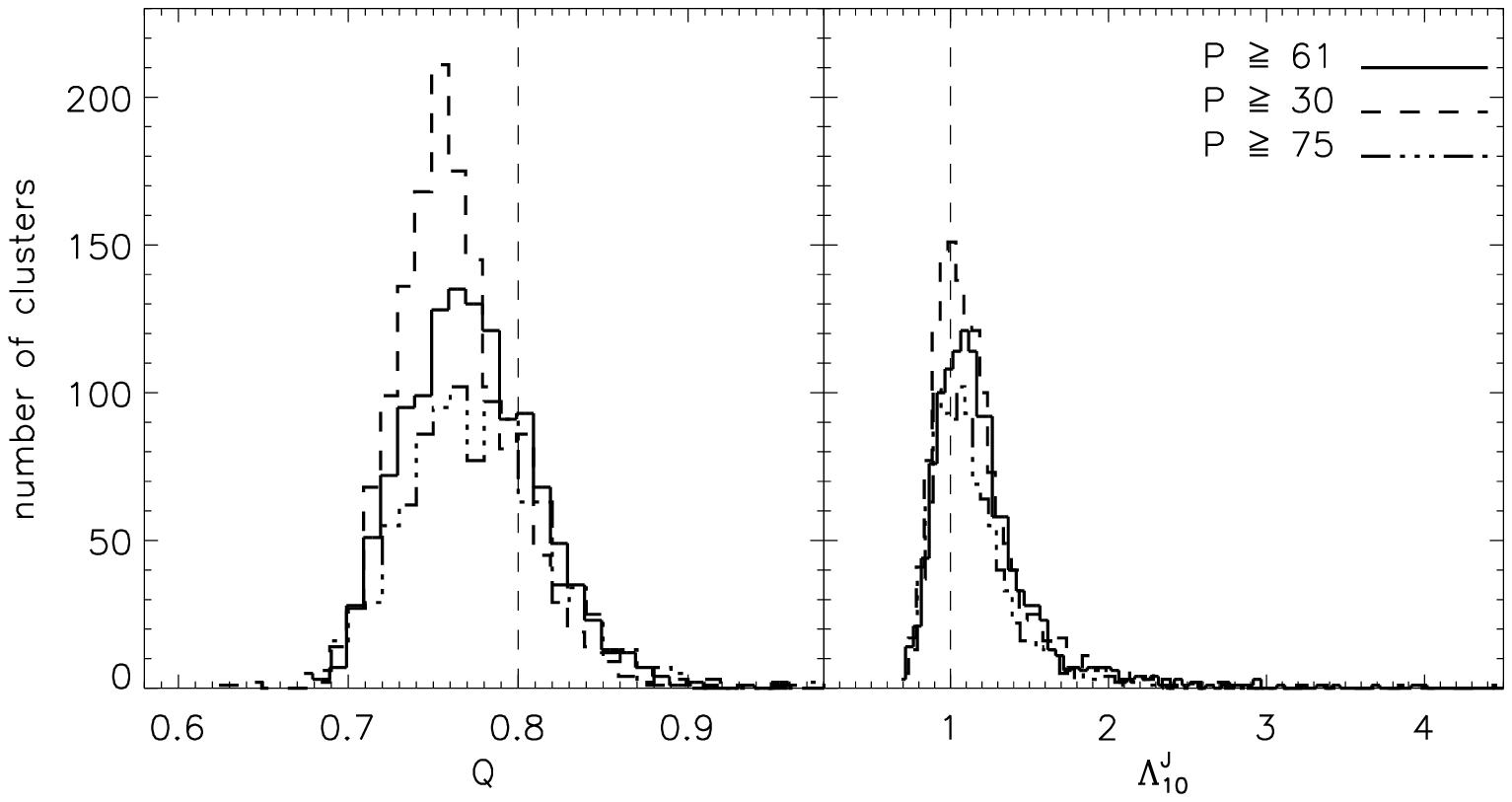,width=\columnwidth}
\end{center}
\vspace{0.5cm}
\caption{The dependence of the distribution of the structure parameter \Q\ (left panel) and of the mass segregation ratios $\Lambda^{J}_{10}$ (right panel) on the choice of the stellar membership probability, $P$. The fiducial case corresponds to the case with $P \geq 61\%$ (full line, 1276 clusters). A higher/smaller value of the threshold probability reduces/increases the number of clusters in the sample. For $P \geq 30\%$ and $P \geq 75\%$, the number of clusters is $1464$ and $998$, respectively. The vertical dotted lines show the division between hierarchical and centrally concentrated clusters at $\Q\ = 0.8$ and the division between non-mass-segregated and mass-segregated clusters at $\Lambda^{J}_{10}=1$, respectively.}
\label{fig2app}
\end{figure}

Here, we compare the distributions of \Q\ and $\Lambda_{10}^{J}$ obtained with different value of the number of most massive stars considered for mass segregation, $n_{\rm MST}$, and of the stellar membership probability, $P$, to those of the fiducial case where $n_{\rm MST}=10$ and $P \geq 61\%$. Fig.~\ref{fig1app} displays the distribution of \Q\ (left panel) and of $\Lambda_{10}^{J}$ for $n_{\rm MST}=5,10$ and $15$, with $P$ being fixed at $P \geq 61\%$. The total number of clusters is the same (i.e., 1276 clusters). The distributions of \Q\ are identical. For higher $n_{\rm MST}$ values, the peak of the distribution of $\Lambda_{10}^{J}$ is shifted towards $\approx 1$. This is not too surprising since for higher value of $n_{\rm MST}$, the distribution of the $n_{\rm MST}$ most massive stars becomes more similar to the one of the total stellar population in the cluster. The arithmetic mean values and standard deviations are $[0.778\pm 0.039,0.778\pm 0.039,0.778\pm 0.039]$ and $[1.46\pm 0.73,1.23\pm 0.37,1.14\pm 0.25]$, whereas the median values of the \Q\ and $\Lambda_{10}^{J}$ distributions are $[0.776,0.776,0.776]$ and $[1.27,1.15,1.08]$ for $n_{\rm MST}=5,10$, and $15$, respectively.

For the cases with different membership probability $P$, a higher value of the threshold probability reduces the number of clusters that fulfil our selection criterion of $N_{*} \geq 40$, and the reverse is true for  smaller $P$ values. Fig.~\ref{fig2app} displays the distributions of \Q\ (left panel) and of $\Lambda_{10}^{J}$ (right panel) for cases with $P \geq 30\%$ (1464 clusters), $P \geq 61\%$ (fiducial, 1276 clusters), and $P \geq 75\%$ (998 clusters). In all three cases here, $n_{\rm MST}=10$. The effect of low $P$ is to contaminate the clusters with mostly low mass field stars. In turn, this shifts the peak of the $\lambda_{10}^{J}$ distribution towards lower values, and it also causes the structure of the clusters to be less centrally condensed (i.e., smaller \Q\ values). The arithmetic mean values and standard deviations are $[0.771\pm 0.078,0.778\pm 0.039,0.780\pm 0.042]$ and $[1.20\pm 0.34,1.23\pm 0.37, 1.18\pm 0.29]$, whereas the median values of the \Q\ and $\Lambda_{10}^{J}$ distributions are $[0.765,0.776,0.777]$ and $[1.12,1.15,1.11]$ for $P \geq 30\%$, $\geq 61\%$ and $\geq 75\%$, respectively.

\end{document}